# Personality-adapted multimodal dialogue system

Tamotsu Miyama[1], Shogo Okada[1]
tamotsu.miyama@jaist.ac.jp, okada-s@jaist.ac.jp
[1]*Japan Advanced Institute of Science and Technology*

*Abstract*— This paper describes a personality-adaptive multimodal dialogue system developed for the Dialogue Robot Competition 2022. To realize a dialogue system that adapts the dialogue strategy to individual users, it is necessary to consider the user's nonverbal information and personality. In this competition, we built a prototype of a user-adaptive dialogue system that estimates user personality during dialogue. Pretrained DNN models are used to estimate user personalities annotated as Big Five scores. This trained model is embedded in a dialogue system to estimate user personality from face images during the dialogue. We proposed a method for dialogue management that changed the dialogue flow based on the estimated personality characteristics and confirmed that the system works in a real environment in the preliminary round of the Dialogue Robot Competition 2022. Furthermore, we implemented specific modules to enhance the multimodal dialogue experience of the user, including personality assessment, controlling facial expressions and movements of the android, and dialogue management to explain the attractiveness of sightseeing spots. The aim of dialogue based on personality assessment is to reduce the nervousness of users, and it acts as an ice breaker. The android's facial expressions and movements are necessary for a more natural android conversation. Since the task of this competition was to promote the appeal of sightseeing spots and to recommend an appropriate sightseeing spot, the dialogue process for how to explain the attractiveness of the spot is important. All results of the subjective evaluation by users were better than those of the baseline and other systems developed for this competition. The proposed dialogue system ranked first in both "Impression Rating" and "Effectiveness of Android Recommendations". According to the total evaluation in the competition, the proposed system was ranked first overall.

## I. INTRODUCTION

In recent years, dialogue systems such as communication robots and voice chats have spread in society. Toward implementing dialogue systems that naturally converse like humans, systems need to understand spoken language and nonverbal information observed by human users, including facial expressions and acoustic features. From this background, many studies have focused on developing multimodal dialogue systems that adapt dialogue strategies based on multimodal inputs [1,2].

It is useful to estimate the personality traits of humans by considering nonverbal information such as facial expressions and prosody [3]. Generally, humans can choose the appropriate words depending on the other person's personality and tailor their speech to the other person. This is especially important for people in the hospitality industry, counselors, travel agents, and other professionals who often interact with people. Dialogue systems should adapt the dialogue strategy based on user attributes such as personality.

Researching user personalities for developing dialogue systems, Sato et al. [8] investigated the relevance of user psychological characteristics in person-to-person dialogue, and Yamamoto et al. [7] investigated the relationship between personality traits in conversations between a user and a dialogue system. In these studies, psychological scales studied in psychology, such as the Big Five [9], were used to measure user personality. However, most of these studies have only conducted personality relationship surveys and have not realized a dialogue system that estimates user personality during the dialogue and adapts to the user.

In this study, we proposed a prototype of a multimodal dialogue system that estimates personality traits during conversation and adapts to the user. Following related studies, we used scores of the Big Five as a user's personality trait to measure personality, which is widely adopted in psychological research. The first step in developing the system is to train a model for estimating user personality using the DNN model. Next, the estimation model is integrated into the dialogue system to estimate user personality during the dialogue. The system then changes the questions from the next turn according to the estimated personality traits. Specifically, based on the estimated extraversion level, the system asks positive questions to users with high extraversion and somewhat less positive questions to users with low extraversion. The personality assessment function was implemented using a personality trait estimation model trained with a machine learning technique. The aim of personality assessment by a robot is to entertain users as an icebreaker [12] in a dialogue system.

Furthermore, because the evaluation metrics regarding the "Effectiveness of Android Recommendations" are important in this competition, we developed a dialogue management module to appeal to the sightseeing spots. We participated in the preliminary round of the Dialogue Robot Competition 2022 [4,5], and the results of the system evaluation by users showed that the proposed personality-adaptive multimodal dialogue system obtained better evaluation performance than other systems.

## II. RELATED WORKS

Research on dialogue systems related to personalities and characters includes research on dialogue systems with specific characteristics and user-adaptive dialogue systems.

Sato et al. [8] analyzed the ease of talking using each psychological scale by conducting human-to-human dialogue experiments. They analyzed both the "ease of talking" and "difficulty of talking" per psychological trait. They used four psychological scales, private speaking tendency, cognitive need, social speaking tendency, and trait shyness, and measured the psychological traits of the experimental participants through questionnaires. Using the measured psychometric traits, they investigated the relationship between speaking ease and speaking difficulty by psychometric traits on dyadic human-human conversation data.

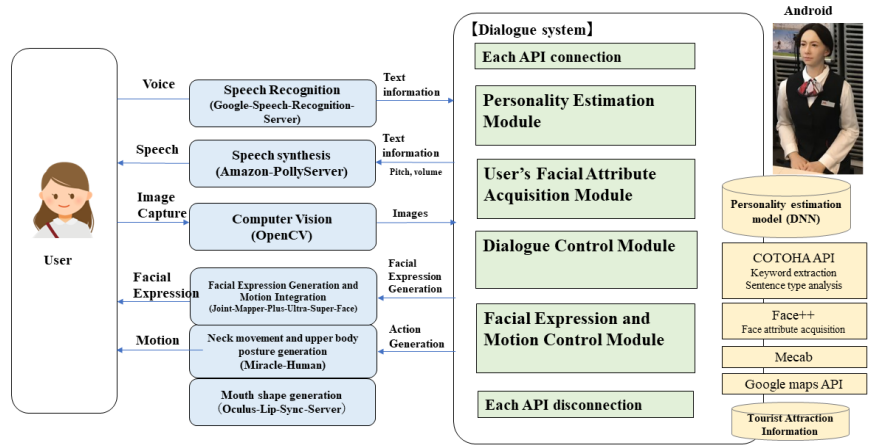

Figure 1: Overview of the dialogue system

Yamamoto et al. [7] investigated the relationship between user personality and the character of a dialogue system to realize a dialogue system that adapts to the user's personality. They used a four-class classification of the Big Five scale and conducted a human–android dialogue experiment using the Wizard-of-Oz (WOZ) method for analyzing the impression and compatibility of the system per specific personality trait class. The personality trait classes are role model, reserved, self-centered, and introvert, which are clustered on the Big Five scale. They investigated impressions and compatibility by conducting a human-to-android dialogue experiment.

Guo et al. [6] investigated user personality adaptability in a task-based dialogue system by focusing on the success or failure of dialogue in a task-oriented, chat-type, multidomain dialogue system controlled with Wizard-of-Oz by recognizing personality traits through questionnaires.

Pierre Y et al. [13] investigated how the personality of a dialogue system influences the persuasiveness of the dialogue system. The study implemented a dialogue system with a speech synthesis model based on personality type.

Itoh et al. [15] investigated the compatibility between users and anthropomorphic agents using anthropomorphic agents with personalities. They used egograms for personality assignment. An egogram is a personality assessment method based on the ego state in transactional analysis that classifies the human mind into five categories. They investigated the subjects' personalities by a questionnaire and investigated their compatibility by conducting a conversation experiment with anthropomorphic agents of different personalities.

Although most of these studies focused on analyzing impressions and compatibility per personality type, ease of speaking per character of the system, and whether the task was successful, they did not implement a personality trait recognition model working in dialogue systems during conversations. Furthermore, other studies did not investigate the effectiveness of personality trait recognition of dialogue systems. Therefore, in this study, we developed a multimodal dialogue system that estimates user personality during the dialogue and adapts the dialogue strategy (selecting appropriate questions based on the personality trait type). We believe that developing this prototype system based on personality trait assessment and adaptation is the first step toward implementing a personality-adapted multimodal dialogue system.

### III. Dialogue System Overview

The overall configuration of the dialogue system is shown in Figure 1. This dialogue system is based on one-to-one dialogue between the android and the user. It achieves dialogue functions in conjunction with various software that controls the android. The dialogue system has the following four modules.

- Personality estimation module
- User's facial attribute acquisition module
- Dialogue control module
- Facial expression and motion control module

The various software include the following, which acquire the user's voice and control the android's speech, facial expressions, and movements.

- Speech recognition

    (Google-speech-recognition-server)

- Speech synthesis (Amazon-Polly-server)
- Computer vision (OpenCV)
- Facial expression generation and motion integration

    (Joint-mapper-plus-ultra-super-face)

- Neck motion upper body posture generation

    (Miracle-human)

- Mouth shape generation

    (Oculus-lip-sync-server)

Each module of the dialogue system is also connected to the DNN pretrained personality trait recognition model and Face++ to acquire the user's personality traits and facial attributes. Furthermore, these modules use the COTOHA API for keyword extraction, Google Maps API for acquiring information on nearby restaurants and cafes, Mecab for the analysis of recognized text information, and, since the task of this Dialogue Robot Competition is a travel agency task, it also reads information on sightseeing spots related to this task. The sightseeing spot data contain six sightseeing spots, which are the data needed for this task.

### A. Personality Estimation Module

First, a DNN pretrained model for personality estimation is loaded, face images are captured using a webcam, and personality traits of the Big Five are estimated from the face images through the model. The estimated personality traits are transferred to the dialogue control module. The machine learning procedure of the personality trait recognition model is described in detail in IV-A.

### B. User's Facial Attribute Acquisition Module

After connecting to the Face++ API, this module extracted various attributes of the user's face from the captured image. These facial features are transferred to the dialogue control module.

### C. Dialogue Control Module

The dialogue control module manages the entire dialogue flow. This dialogue control module interacts with speech recognition software (This "Google-speech-recognition-server" software.) and speech synthesis software (This "Amazon-Polly-server" software.) and controls the dialogue using the COTOHA API, sightseeing spot data, and Google Maps API. After speech recognition, the system performs keyword extraction, example retrieval, and speech generation using information converted to text by Google-speech-recognition-server. Additionally, it considers pitch, volume, and turn-taking. The system also acquires information on sightseeing spots and user-selected sightseeing spot IDs (CompetitionID-Server) to generate speech related to sightseeing spots.

### D. Facial Expression and Motion Control Module

This module receives speech information generated by the dialogue control module and connects to facial expression generation and motion integration software (called "joint-mapper-plus-ultra-super-face" software), neck motion upper body posture generation software (called "miracle-human" software), and mouth shape generation software (called "oculus-lip-sync-server" software) to create facial expressions and postures for the android.

## IV. PERSONALITY TRAIT RECOGNITION MODEL

We used the Big Five scale to estimate personality. The Big Five is a taxonomy of personality traits widely used in psychological research. It is based on the theory that human personality can be explained by five factors: extraversion, agreeableness, conscientiousness, neuroticism, and openness.

### A. Training model

We used the First Impressions V2 dataset [11], which contains approximately 10000 videos and Big Five scores annotated by third-party coders. In general, Big Five scores are determined using the results of questionnaire responses provided by the participant themselves, which measures personality, but in this dataset, third-party coders annotated the Big Five scores of participants (people in videos) to annotate their impression of what kind of personality they appear to have.

| dataset | First Impressions V2 |
|---|---|
| preprocessing | We extracted data with high (low) personality traits from the First Impressions V2 video data and represented them as still images using OpenCV. |
| data | 2500 face images (250 images for each high (low) personality trait) 【details】 Extraversion high/low: 500 (train: 400, val: 100) Agreeableness high/low: 500 (train: 400, val: 100) Conscientiousness high/low: 500 (train: 400, val: 100) Neuroticism high/low: 500 (train: 400, val: 100) Openness high/low: 500 (train: 400, val: 100) |
| Machine Learning Model | Transfer learning with pretrained VGG16, binary classification |
| Usage environment, Framework | GoogleColab, PyTorch |

TABLE II. USER PERSONALITY ESTIMATION MODEL

|  | *Loss* | *Acc* |
|---|---|---|
| Extraversion | 0.3652 | 0.8300 |
| Agreeableness | 0.4714 | 0.7300 |
| Conscientiousness | 0.5463 | 0.7100 |
| Neuroticism | 0.3773 | 0.8265 |
| Openness | 0.4911 | 0.7800 |

TABLE III. recognition accuracy of personality traits

To perform real-time processing, a simple personality estimation model is trained using single images (not using video data). As a preprocessing step, we extracted data from the First Impressions V2 video data in which each personality trait value was represented in the face image and then used OpenCV to extract the still images. Specifically, to capture still images with high extroversion, First Impressions V2 video data were extracted in order of extroversion and represented as still images. Two hundred images with high/low personality scores for each trait are used as training data (2000 images: 400 * five traits).

### B. Neural Network architecture and Evaluation

The model was finetuned using training images by transfer training with a large-scale model [13], which was pretrained. The DNN architecture used was VGG16, a convolutional neural network (CNN) consisting of 16 layers. Training was performed for each of the five personality traits, and two labels were predicted for each personality trait, one high and one low. The batch size was 32, the number of epochs was 8, the loss function was a cross-entropy error, the optimization method was SGD, and the learning rate was 0.001.

The evaluation results are shown in TABLE III. The trained model's recognition accuracy (Acc) was more than 0.71 (71%). Although the accuracy should be improved, the accuracy is moderate for a binary classification task. The extraversion recognition model was used for dialogue management in the proposed system because the result is best in all traits (0.83). The model was incorporated into a multimodal dialogue system.

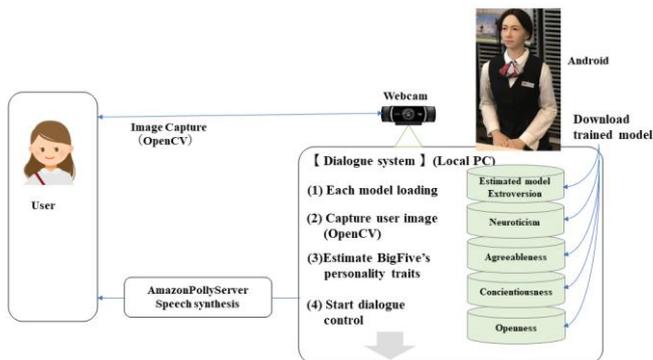

Figure 2: Dialogue system incorporating the estimation model

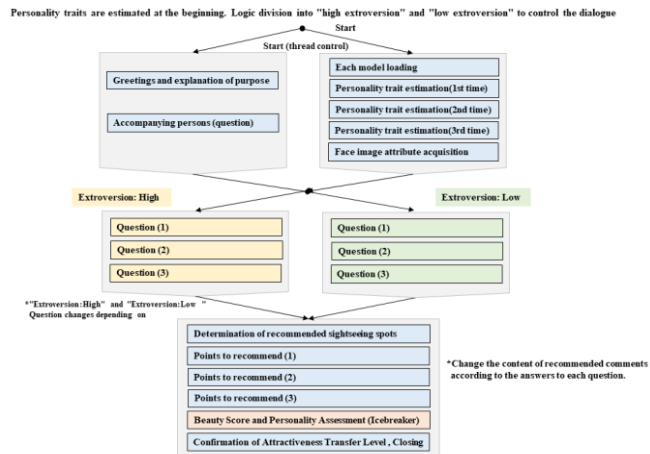

Figure 3: Dialogue flow

## V. DIALOGUE FLOW

Figure 2 shows an overview of the proposed dialogue system with a personality trait recognition model. The trained personality trait recognition model is incorporated into the dialogue system to enable real-time user personality recognition. Face images were captured three times at the beginning of the conversation, personality traits were estimated for each of the three times, and the final personality traits were classified by the average of the three output values.

The dialogue flow of the dialogue system for this competition is shown in Figure 3. The Big Five have several personality scales, but this time, we focused on extroversion to compose the dialogue flow. Since the recognition of a personality takes some time, the conversation process and the personality traits estimation process were executed in parallel at the start of the dialogue flow. After the personality traits were estimated, the dialogue flow was divided into two parts, one for high extraversion and the other for low extraversion, and the subsequent dialogue flow was executed. The questions were varied according to high and low extraversion, and the recommended comments were varied according to the answers to each question. The system asked positive questions to users with high extraversion and fewer positive questions to users with low extraversion.

## VI. CONTROL OF THE MULTIMODAL SYSTEM

This multimodal dialogue system needs to be controlled to increase the satisfaction of the conversation with the user and the evaluation criteria of the task.

### A. Personality assessment

Using the estimated result of the Big Five personality traits, the system conducted a "personality assessment" of the user using a trained model in Section IV. The personality assessment is an ice-breaker function and relieves tension in a conversation with the android. Since the system was already able to recognize the user's personality traits from their face image in the dialogue system, personality assessment was performed by commenting on the positive aspects of each personality trait.

### B. Appealing the attractiveness of sightseeing spots

In this competition, one of the evaluation metrics was the "Effectiveness of Android Recommendations. For this evaluation metric, this dialogue process is designed to appeal to the recommended points of sightseeing spots multiple times. As described in the dialogue flow, the system decided to comment on three recommendation points for the recommended sightseeing spots and fully convey the attraction of the sightseeing spots. Furthermore, minor controls were made, such as raising the volume of the voice during the part where the recommended points were uttered.

### C. Facial expressions and movements of androids

Since travel agencies are customer service businesses, the androids were designed to speak with a basic smile. After the user utters, the android always responds with its voice, facial expressions, and actions. We aims to notice the user that the android is aware of what the user has said, and generated a somewhat over-the-top nodding behavior. The android's face was tilted to the right when explaining the photos of the sightseeing spots on the right. The behavior promotes the joint attention of the user to the photos. The android's movements and facial expressions were manually designed so that they would interact with the user with human-like motion. Developing an automatic generation model of natural nonverbal behavior is future work.

## VII. REAL-WORLD VERIFICATION AND VALIDATION RESULTS

We tested our personality-adaptive multimodal dialogue system in a real-world environment in the preliminary round of the 2022 Dialogue Robot Competition. We decided to use this competition as an experimental environment because it allows us to receive evaluations from general users. We compared our evaluation to the baseline system created by the competition organizers and to the average evaluations of the other teams.

| | Our team | Baseline | Average of other teams |
|---|---|---|---|
| satisfaction with choice | 5.38 | 4.19 | 4.40 |
| sufficiency of information | 5.46 | 3.96 | 4.38 |
| naturalness of dialogue | 5.35 | 3.81 | 3.60 |
| appropriateness of dialogue | **5.35** | 4.41 | 4.24 |
| likability of dialogue | **5.46** | 4.59 | 4.53 |
| satisfaction with response | **5.62** | 4.15 | 4.35 |
| trust in the other party | **4.92** | 4.30 | 4.30 |
| helpfulness of information | 5.73 | 4.67 | 4.76 |
| intention to reuse | 5.19 | 4.07 | 4.15 |
| **Impression Total** | 48.46 | 38.15 | 38.72 |

TABLE III. Evaluation Results (Impression)

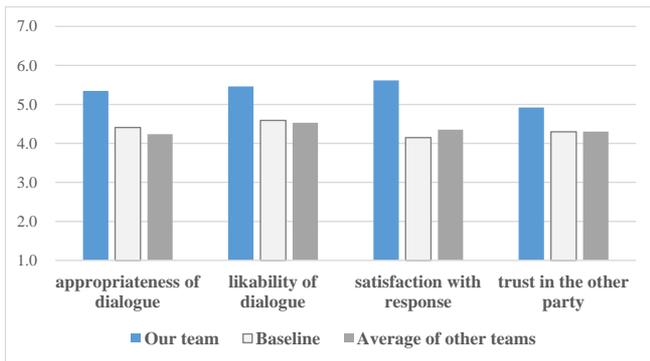

Figure 4: Evaluation Results (Impressions)

### A. Competition Tasks and Evaluation Metrics

The task of this competition was a travel agency task. The user preselects two sightseeing spots from a list, including six sightseeing spots provided by the competition organizers, and the android recommends one of the destinations. System evaluation was based on a 50% rating of whether the user was more interested in visiting the recommended sightseeing spot than before the conversation and a 50% impression rating of the user's satisfaction. The effectiveness of the android's recommendations was evaluated by the difference between scores of whether the user wants to visit the recommended sightseeing spot before and after the conversation. The impression ratings included "satisfaction with choice," "sufficiency of information," "naturalness of dialogue," "appropriateness of response," "trust in the other party," "helpfulness of information," and "intention to reuse," each of which is rated on a 7-point scale from 1 to 7. The items are rated on a 7-point scale from 1 to 7. Our team had 26 participants, men and women, from their teens to their seventies. There were a total of 13 teams, including the baseline team.

### B. Evaluation Results

The results of the evaluation of impressions at the preliminary round are shown in Table III and Figure 4. The evaluation items related to personality adaptation, "appropriateness of dialogue," "likability of dialogue," "satisfaction with a response," and "trust in the other party", were all significantly higher than the those of the baseline and average results of the other teams.

| | Impression Rating Total | Android Recommendation Effect |
|---|---|---|
| Our team | **48.46** | **45.88** |
| Baseline | 38.15 | 5.38 |
| Average of other teams | 38.71 | 10.36 |

TABLE IV. Evaluation Results (Overall)

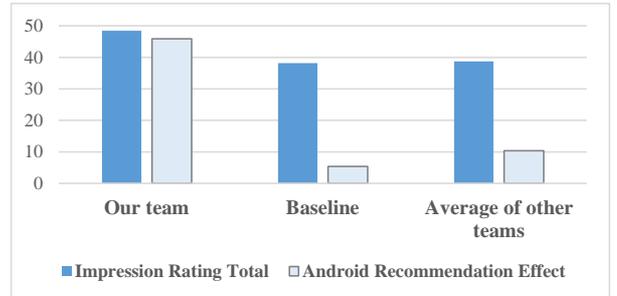

Figure 5: Evaluation Results (Overall)

The evaluation results for the competition are shown in Table IV and Figure 5. Our team's "Impression Rating" and "Effectiveness of Android Recommendations" were significantly higher than those of the baseline and the average of the other teams, making us the first-place team overall in the evaluation results.

In particular, the effectiveness of the android's recommendation was evaluated to be much higher than the baseline and the average of other systems. There is a possibility that this is because the team's dialogue system sufficiently promoted the attractiveness of the recommended sightseeing spots. This allowed the team to meet the task evaluation metrics required in the competition, leading to a high overall evaluation.

We consider personality assessment. The android's facial expressions and movements also contributed to the score.

## VIII. CONCLUSION

This paper describes a multimodal dialogue system that adapts a dialogue strategy based on the recognition results of personality for user adaptation. For personality adaptation in dialogue systems, we propose a dialogue system that estimates user personality traits during the dialogue and automatically changes types of questions based on the estimated results.

To estimate personality traits during the dialogue, we trained a DNN model and built a dialogue system incorporating the estimated model. Furthermore, we experimented with this system in the preliminary round of the 2022 Dialogue Robot Competition. The results showed that the system was highly rates in terms of the evaluation items related to personality adaptation.

A direction of further research is to investigate the relationship between user personality and dialogue system personality. Furthermore, due to the nature of the experimental environment of competition, the dialogue system was evaluated based on the criteria of tailored tasks of the competition, and it is necessary to conduct further research and experiments in personality adaptive dialogue systems.

In addition to personality and multimodal information, various factors need to be considered to realize a dialogue system using a naturally speaking android. We need to integrate our knowledge with research on speech recognition, text analysis, dialogue control, turn-taking, pitch, voice volume and speed, dialogue development, android facial expression generation, android motion generation, and many other factors.

## APPENDIX

| sightseeing spots Category | Extraversion | Question (1) | Question (2) | Question (3) |
|---|---|---|---|---|
| Museums/galleries Towers and Observation facilities | High | Indoor or Outdoor | Do you like sports? Confirm name of sport | History or Art Movie or Music |
| | Low | Indoor or Outdoor | Sweet or Spicy Confirm name of sweets | History or Art Movie or Music |
| Museums/Science Museums/Resources and Parks | High | Indoor or Outdoor | Do you like sports? Confirm name of sport | Sweet or Spicy History or Art |
| | Low | Indoor or Outdoor | Means of transportation | Sweet or Spicy Sweets Name Confirmation History or Art |

The system asked positive questions to users with high extraversion (e.g., sports) and fewer positive questions to users with low extraversion (e.g., means of transportation,). Questions varied according to extroversion level and sightseeing spot category.

TABLE V. EXAMPLE QUESTION.

## ACKNOWLEDGMENT

We would like to thank the Organizers of the Dialogue Robot Competition 2022 for providing the necessary software, androids, experimental environment, and various opportunities for academic research to build this dialogue system.